\documentclass{PoS}

\usepackage{exscale}                  
\usepackage[intlimits]{amsmath}       
\usepackage{amsfonts}
\usepackage{amssymb,amscd}
\usepackage[dvips]{epsfig}                   
\usepackage{array}
\usepackage{wrapfig}
\usepackage{bbm}
\usepackage{multirow}

\newcommand{\beqa}{\begin{eqnarray}}
\newcommand{\eeqa}{\end{eqnarray}}
\newcommand{\beq}{\begin{equation}}
\newcommand{\eeq}{\end{equation}}
\newcommand{\pslash}{p\hspace{-.5em}/\hspace{.15em}}

\newcommand{\Pslash}{P\hspace{-.55em}/\hspace{.20em}}

\newcommand{\tr}{\textrm{tr}}

\newcommand{\lqcdsq}{\Lambda^2_{\mathrm{QCD}}}

\title{Pion-cloud effects in the BSE description of mesons}

\ShortTitle{Pion-cloud effects in the BSE description of mesons}

\author{\speaker{Richard Williams}$^{a}$\\
Institut f\"ur Physik, TU Darmstadt,
	Schlossgartenstr. 9,
	D-64289 Darmstadt, Germany\\
E-mail: \email{richard.williams@physik.tu-darmstadt.de}}
\author{Christian S. Fischer$^{ab}$\\
	 \llap{$^a$} Institut f\"ur Physik, TU Darmstadt,
	Schlossgartenstr. 9,
	D-64289 Darmstadt, Germany \\
\llap{$^b$}GSI Helmholtzzentrum f\"ur Schwerionenforschung GmbH,  Planckstr. 1,
64291 Darmstadt, Germany. }

\abstract{
We investigate the effect of including hadronic resonance contributions
in the description of light quarks and mesons. To this end we take into account the 
back-coupling of the pion onto the quark propagator within 
the non-perturbative continuum framework of 
Schwinger-Dyson equations (SDE) and Bethe-Salpeter equations (BSE), in
essence describing the so-called \emph{pion-cloud}. 
As a result of our study we find that an unquenching of this form
provides for considerable effects in the spectrum of light mesons.
}

\FullConference{8th Conference Quark Confinement and the Hadron
Spectrum\\September 1--6, 2008\\Mainz, Germany}

\begin{document}

\section{Introduction}\label{sec:intro}
In QCD it is not the quarks and gluons that are asymptotic states,
directly observable in detectors, but rather colourless composites such
as mesons and baryons. This entails the need to describe in detail the
properties of our final states in terms of their constituent particles,
an inherently complicated task due to the non-perturbative effects of
confinement and dynamical chiral symmetry breaking.

The natural framework for this composite description of mesons in the continuum are the
Schwinger-Dyson and Bethe-Salpeter equations (SDEs and BSEs). Such
studies have been extensively performed in the Rainbow-Ladder
approximation~\cite{Maris:2003vk}, which generally reduce to a quenched study with
non-perturbative effects subsumed into an effective gluonic interaction.
While there have been several studies attempting to improve upon this
simplest of all truncations, in the form of including unquenching
effects due to quarks, \emph{e.g.}~\cite{Watson:2004jq,Watson:2004kd,Fischer:2005en}, we here take a different viewpoint. Instead of
modelling the contribution from quarks directly, we rather consider the dominant
contributions arising from the resultant \emph{mesonic} degrees of freedom. Thus in this
talk we focus upon pion-cloud contributions~\cite{Thomas:2008bd} to the light meson spectrum
in a beyond the rainbow truncation scheme. 

\section{Including the Pion-Cloud}\label{sec2}
\begin{figure}[b]
  \centerline{\includegraphics*[width=0.80\columnwidth]{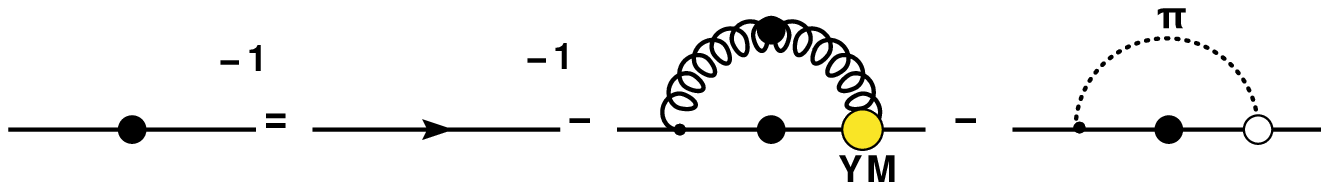}}\vspace{3mm}
\centerline{\includegraphics*[width=0.75\columnwidth]{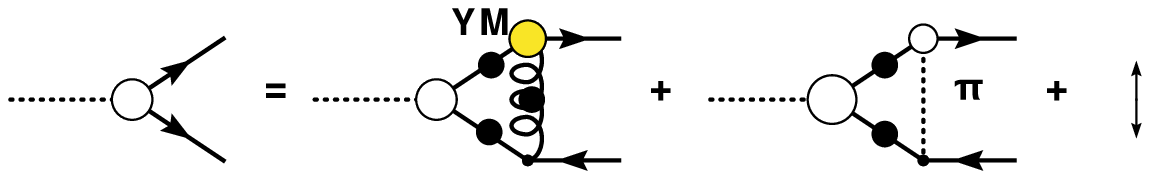}}
\caption{The approximated quark-SDE with effective one-gluon and 
one-pion exchange, together with the corresponding BSE. The up-down arrow
indicates an averaging procedure of the pion-exchange diagram with respect
to the dressed/undressed quark-pion vertex.
}\label{fig:truncation}
\end{figure}

The prescription for including pion degrees of freedom in the SDEs and
BSEs in a manner consistent with the axial-vector Ward-Takahashi identity 
(avWTI) was first presented in~\cite{Fischer:2007ze} and investigated
within
the real-value approximation. Further modifications were suggested 
in~\cite{Fischer:2008sp}, with the resultant system of equations depicted
in Fig.~\ref{fig:truncation}. The first loop diagram of both pictorial
equations relates to the usual rainbow-ladder, where the infrared
suppressed gluon is
enhanced by the vertex dressing, indicated by \emph{YM}. To satisfy the
avWTI this enhancement is restricted to depend on the same momentum as
the exchanged gluon; the two dressings are often combined into an
effective gluon dressing which is consequently modelled.
The second diagram that contributes to the quark-SDE and meson-BSE
represents the back-reaction of the pion onto the quark, \emph{i.e.}
pion-cloud effects. This requires knowledge of the
quark-pion vertex, which we parameterise in terms of the pion
Bethe-Salpeter amplitude. Necessarily this couples the two equations
(whose form is presented in~\cite{Fischer:2008sp,Fischer:2008wy}), forming a highly non-trivial system of integral equations.

\subsection{Yang-Mills Part}
We need to specify the details of our gluon exchange and quark-gluon
vertex that model the interaction. We choose two different
model ans\"atze that have been employed already in previous works.
The first model takes fits to the gluon as obtained from numerical
solutions of the corresponding Schwinger-Dyson
equations. This is taken in combination with results inspired from
solutions of the quark-gluon vertex for a restricted kinematic
section~\cite{Alkofer:2008tt}. This has the virtue that it also provides for generation
of a topological charge and hence an anomalous mass contribution to the
$\eta$, $\eta'$~\cite{Alkofer:2008et}.

Thus, in choosing a particular kinematic configuration for our vertex
dressing we employ a rainbow-ladder form of the interaction (thus
preserving the avWTI), 
\emph{i.e.} 
\begin{equation}
\Gamma_\nu^{YM}(p_1,p_2,p_3) \;=\; \gamma_\nu\, 
{Z_2}/{\widetilde{Z}_3} \, \, \Gamma^{YM}(p_3^2) \,, \label{v1}
\end{equation}
with quark momenta by $p_1$ and $p_2$ and the gluon momentum $p_3$. 
Note, however, that (\ref{v1}) involves only the
$\gamma_\nu$-part of the full tensor structure of the vertex. It has 
been shown in the analysis of the full quark-gluon 
vertex of Ref.~\cite{Alkofer:2006gz} that such a model cannot capture 
all essentials of dynamical chiral symmetry breaking. We refer to this
as the soft-divergent model interaction.

The second model that we consider is that of Maris-Tandy~\cite{Maris:1999nt}. In this
case, the dressing of both the gluon and the vertex is modelled by a
phenomenological ansatz that includes the correct one-loop UV
running and provides dynamical chrial symmetry breaking. We include a study of
this much used model in the context of pion unquenching for comparison.

\subsection{Pionic Contribtion}
The decomposition of a Bethe-Salpeter vertex function $\Gamma(p;P)^{(\mu)}$
is well-established in the literature, with its form constrained by transformation 
properties under CPT~\cite{LlewellynSmith:1969az}. In particular the pion is 
given by the following form
\begin{eqnarray}
\Gamma^j_\pi(p;P)\!\!&=&\!\!\!\! \tau^j\gamma_{5}\Big[F_1(p;P)
-i\Pslash F_2(p;P)\label{pion}-i\pslash \left(p\cdot P\right)F_3(p;P)
-\left[\Pslash,\pslash\right]F_4(p;P)\Big]\, .
\end{eqnarray}
This is of particular relevance for our pionic part of the interaction.
We approximate the full pion Bethe-Salpeter wave function in the quark-SDE 
and the kernel of the BSE by the leading amplitude in the chiral limit given
by
\begin{equation}
\Gamma^j_{\pi}(p;P) = \tau^j \gamma_5 \frac{B_\chi(p^2)}{f_\pi}\;. \label{piapprox}
\end{equation}
Here $B_\chi(p^2)$ is the scalar dressing function of the quark propagator
in the chiral limit. The effects of neglecting the three sub-leading 
amplitudes have been quantified for a real-value approximation in 
Ref.~\cite{Fischer:2007ze}. The great advantage of the approximation (\ref{piapprox}) compared to the
full back-coupling performed in Ref.~\cite{Fischer:2007ze} is that we can
then fully take into account the quark propagator in the complex plane as
necessary input into the Bethe-Salpeter equation. Whilst proving to be a
simple prescription in itself, it gives rise to the technical challenge
of having to evaluate the full normalisation condition of the BSE due to
the non-trivial momentum dependence of the exchange kernel. We discuss
this in the next section.

\section{Normalisation \label{res}}
Since we solve homogeneous equations for the Bethe-Salpeter amplitude,
their subsequent renormalisation comes as an auxiliary condition derived
from the inhomogeneous BSE:
\begin{eqnarray}\label{norm}
\delta^{ij}&=&2\frac{\partial}{\partial P^2} \tr\int
\frac{d^4k}{(2\pi)^4}\Bigg[3 \, \, \bigg(
\overline{\Gamma}_\pi^i(k,-Q)  S(k+P/2)\Gamma_\pi^j(k,Q)S(k-P/2)
\bigg) \nonumber\\
&&\hspace{2.20cm}+\int \frac{d^4q}{(2\pi)^4} 
[\overline{\chi}_\pi^i]_{sr}(q,-Q)
K_{tu;rs}^{\textrm{pion}}(q,k;P)[{\chi}_\pi^j]_{ut}(k,Q)\Bigg]\;,\nonumber
\end{eqnarray}
where $Q^2=-M^2$ is fixed to the on-shell meson mass,
the trace is over Dirac matrices and the 
Bethe-Salpeter wave-function $\chi$ is defined by
$\chi_\pi^j(k;P) = S(k+P/2)\Gamma^j_\pi(k,P)S(k-P/2)\,$. The conjugate
vertex function $\bar{\Gamma}$ is given by $\bar{\Gamma}(p,-P)=C
\Gamma^T(-p,-P)C^{-1}$, with the charge conjugation matrix $C=-\gamma_2\gamma_4$.

The first term of (\ref{norm}), independent of the kernel, is easily evaluated since
one needs only derivatives of the quark propagator. The
second term is substantially more complicated due to the double
integration over the interaction kernel with respect to two
four-momenta, $k$ and $q$. We evaluate the
integral of (\ref{norm}) and employ finite difference methods to compute
the derivative. We find that the contribution
from the kernel provides important contributions.

\section{Results}
The parameters of for both model interactions were fit to pion
observables, with the additional constraint of the topological charge
for our soft-divergent interaction. In constraining the parameter set
to meson observables, we find for our model a quark mass
of $m_{\overline{\textrm{MS}}} = 3.4$~MeV at $\mu=2$~GeV, whilst for Maris-Tandy
we have $m_{\overline{\textrm{MS}}} = 4.4$~MeV. The remaining
parameters of the interaction are:
\begin{center}
\begin{tabular}{@{}cccc|cccc}
  \multicolumn{4}{c|}{Soft-Divergent Interaction} &
  \multicolumn{4}{c}{Maris-Tandy Interaction}\\
\hline
  $d_1$ & $d_2$ & $d_3$ & $\lqcdsq$ &  $\omega$ &
  $D$  & $m_t$  & $\lqcdsq$\\
  (GeV$^2$) & (GeV$^2$) & (GeV$^2$) & (GeV$^2$) & (GeV) & (GeV$^2$) &
  (GeV) & (GeV$^2$) \\
\hline\hline
  $1.45$  & $0.1$  & $3.95$ & $0.52$    & 
  $ 0.37$ & $1.45$ & $0.5$  & $0.234^2$ \\
\hline
\end{tabular}
\end{center}
We calculated a range of meson observables, detailed in
Table~\ref{results}. We observe that the effect of the pion
back-reaction has only a small impact on the pion mass itself, resulting
in a small positive or negative shift depending upon the form of the
interaction.  The impact of including pion-cloud effects on the leptonic
decay constant is fairly large, with effects of the order of $10\%$.

For the remaining heavier mesons, the common trend is that the inclusion
of such an unquenching gives rise to negative mass shifts of
$100$--$200$~MeV. Most notable of these are for the rho, where we
predict that unquenching from the pion-cloud yields a bound-state $\sim100$~MeV
lighter than in the quenched theory, in line with recent lattice
simulations~\cite{leinweber}.

\begin{table}[t!]
\centering
\begin{tabular}{@{}cc|cc|ccccc|cc}
  \multicolumn{2}{c|}{Model Employed} & $M_\pi$ & $f_\pi$ & $M_\sigma$ &
  $M_\rho$ & $f_\rho$ & $M_{a_1}$ & $M_{b_1}$ & $M_\eta$ & $M_{\eta'}$\\
\hline
\hline
\multirow{2}{*}{Maris-Tandy} & w/o pi  & 140        & 104 & 746 & 821 & 160 & 979 & 820 & & \\
                             & inc. pi & 138$^\dag$ & 93.2$^\dag$& 598 & 720 & 167 & 913 & 750 & & \\
\hline
\multirow{2}{*}{Our Model}   & w/o pi  & 125        & 102 & 638 & 795 & 159 & 941 & 879 & 493& 949\\
                             & inc. pi & 138$^\dag$ & 93.8$^\dag$& 485 & 703 & 162 & 873 & 806 & 497 & 963 \\
\hline
\hline
Experiment                   &         & 138 & 92.4&400--1200& 776 & 156 &1230 &1230 & 548 & 948 \\
\end{tabular}
\caption{BSE results for a range of mesons in the Maris-Tandy and soft-divergence models
employed with (`incl.') and without (`w/o') the pion back-reaction.
Model parameters are tuned such that  values marked by $\dag$ are
reproduced when the pion-exchange kernel is switched on.  Results for
rainbow-ladder without pion effects are for the same parameter set. 
\label{results}}
\end{table}

It is clear, however, that in order to reproduce the rich spectrum of
light mesons that we need to include spin dependent contributions 
from the Yang-Mills part of the quark-gluon vertex. It is envisaged that
this will indeed have a strong impact on the calculated masses of
bound-states and is the object of future research.

\section*{Acknowledgements}
This work has been supported by the Helmholtz-University 
Young Investigator Grant No. VH-NG-332.



\end{document}